\documentclass[12pt,a4paper]{article}
\usepackage{graphicx}
\usepackage{latexsym}
\usepackage{amsmath}
\usepackage{amsfonts}
\usepackage{amssymb}
\newtheorem{theorem}{Theorem}
\newtheorem{proposition}[theorem]{Proposition}
\newenvironment{proof}[1][Proof]{\textbf{#1.} }{\ \rule{0.5em}{0.5em}}

\begin{document}

\title{Sharing secret color images using cellular automata with
memory}
\author{G. \'{A}lvarez Mara\~{n}\'{o}n, L. Hern\'{a}ndez
Encinas\thanks{Corresponding author. Tel.: (+34) 915 618 806 (ext.
458), Fax: (+34) 914 117 651, Email: luis@iec.csic.es}\\
{\footnotesize Dpt. Tratamiento de la Informaci\'{o}n y
Codificaci\'{o}n, Instituto de F\'{i}sica Aplicada, CSIC}\\
{\footnotesize C/ Serrano 144, 28006-Madrid, Spain}\\
{\footnotesize E-mails: \{gonzalo, luis\}@iec.csic.es}
\and A. Mart\'{i}n del Rey\\
{\footnotesize Dpt. de Matem\'{a}tica Aplicada, E.P.S.,
Universidad de Salamanca}\\
{\footnotesize C/ Santo Tom\'{a}s s/n, 05003-\'{A}vila, Spain}\\
{\footnotesize E-mail: delrey@usal.es}}
\date{}
\maketitle

\begin{abstract}
A $\left(k,n\right)$-threshold scheme based on two-dimensional
memory cellular automata is proposed to share images in a secret
way. This method allows to encode an image into $n$ shared images
so that only qualified subsets of $k$ or more shares can recover
the secret image, but any $k-1$ or fewer of them gain no
information about the original image. The main characteristics of
this new scheme are: each shared image has the same size that the
original one, and the recovered image is exactly the same than the
secret image; \emph{i.e.}, there is no loss of resolution.
\end{abstract}


\section{Introduction}

A \emph{secret sharing scheme} is a method which allows to share a
secret among a set of users in such a way that only qualified
subsets of these users can recover the secret. Consequently, the
basic idea in secret sharing schemes is to divide the secret into
a fixed number of pieces, called \emph{shares} or shadows, which
are distributed among the participants so that the pooled shares
of certain subsets of users allow the reconstruction of the
secret.

Secret sharing schemes were independently introduced by
Shamir~(\cite{Shamir79}) and Blakley~(\cite{Blakley79}), and their
original motivation was to safeguard cryptographic keys from loss.
These schemes also have been widely employed in the construction
of several types of cryptographic protocols (see, for example,
\cite{MOV97}) and consequently, they have many applications in
different areas such as access control, opening a bank vault,
opening a safety deposit box, or even launching of missiles.

The basic example of a secret sharing scheme is the $\left(
k,n\right) $\emph{-threshold scheme} (or $k$\emph{-out-}$n$
\emph{scheme}) for integers $1\leq k\leq n$. In such scheme there
exists a \emph{dealer} (or mutually trusted party) and $n$
participants. The dealer computes $n$ secret shares $S_{i}$,
$0\leq i\leq n-1$, from an initial secret $S$, and securely
distributes them to the users $P_{0},\ldots,P_{n-1}$, in such a
way that any $k$ or more users who pool their shares may easily
recover the original secret $S$, but any group knowing only $k-1$
or fewer shares is unable to recover the secret. In other words,
each group of $k-1$ or fewer shares reveals absolutely no
information about the secret image. Shamir's scheme, which is
based on polynomial interpolation, and Blakley's scheme, based on
the intersection of affine hyperplanes, are examples of $\left(
k,n\right) $-threshold schemes.

Subsequently, Ito \emph{et al.}~(\cite{ISN87}) and Benaloh
\emph{et al.}~(\cite{BenalohLeichter90}) described a more general
situation based on the specification of the subsets of
participants that should be able to determine the secret and the
subsets of participants that should not. These general secret
sharing methods are intimately related to the notion of
\emph{access structure} (see~\cite{ABSS01, TzengHu02}).

For secret sharing schemes, the \emph{information rate for a
particular participant} is the bit-size ratio:
\begin{equation}
R_{p} = \frac{\text{size of the shared secret}}{\text{size of the
participant's share}}.
\label{rate}
\end{equation}
Moreover, the \emph{information rate for a secret sharing scheme}
is the minimum rate over all participants. In this sense, an
\emph{ideal} secret sharing scheme is a scheme in which the size
of the shares given to each participant is equal to the size of
the secret; consequently, for ideal secret sharing schemes the
information rate is $1$. Moreover, secret sharing schemes
satisfying the additional property that unqualified subsets can
gain absolutely no information about the secret are called
\emph{perfect}. For a more detailed description we refer the
reader to~\cite{MOV97, Stinson92, Stinson02}.

Usually, the secret to be shared consists in text data, but also
images can be considered. The first scheme to share images was due
to Naor and Shamir~(\cite{NaorShamir95}) and it is called
\emph{visual cryptography}. It is based on \emph{visual threshold
schemes} $k$ of $n$, \emph{i.e.}, the original image is divided in
$n$ shares. Each of them is photocopied in a transparency and
then, the original image is recovered by superimposing any $k$
transparencies but no less. Moreover, no cryptographic protocol is
used to recover it. Its main feature is the use of human vision
properties in order to recover the original image.

Due to the characteristics of this model, only black \& white
images were suitable to be shared. Nevertheless, in recent years,
a wide variety of new proposals based on visual cryptography have
emerged not only for processing gray-level images
(\cite{LinTsai03, ThienLin02, VerheulTilborg97}), but also for
color images (\cite{CTC99, Hou03, RijmenPreneel96}). In these
visual threshold schemes each pixel of the secret image is
ciphered by means of $h$ subpixels (the \emph{pixel expansion})
for the $n$ shares; consequently, the size of the shared images is
much bigger than the original one. Moreover, another disadvantage
of these schemes is that there is a great contrast loss between
the secret image and the recovered one.

Furthermore, other algorithms for sharing images, not based in the
visual cryptography paradigm, have appeared (see, for example,
\cite{ChangHwang98, TCC02}).

In this paper, we propose a new graphic sharing scheme;
\emph{i.e.}, a secret sharing scheme for black \& white (b\&w),
gray-level and color images, by means of two-dimensional memory
cellular automata. The proposed scheme is based on cellular
automata and the properties of these kind of discrete dynamical
systems permit us to define an algorithm for sharing secret
images. In the scheme, the shares obtained for each participant
have the same size than the secret image and the recovered image
is exactly the same than the original one, without loss of
resolution. These properties are not satisfied by any other
graphic scheme.

Roughly speaking, two-dimensional cellular automata are time delay
dynamical systems for which time and space are discrete. They
consist of a collection of a finite two-dimensional array of
simple objects, called cells, interacting locally with each other.
Each cell can assume a state such that it changes in every time
step according to a local rule whose variables are the states of
some cells at previous time steps. The prize, compared to visual
cryptography, is that in this new protocol some computations are
needed to recover the original image.

The design of cryptographic protocols by means of two-dimensional
cellular automata is a recent event, and their use are only
restricted to symmetric ciphers for images (see~\cite{AHHMR03,
HMH02}), by the moment. On the other hand, one-dimensional
cellular automata has been widely used not only for symmetric
ciphers: stream ciphers (see, for example \cite{DHHHMRV03,
Wolfram86}) and block ciphers (\cite{Gutowitz93}), but also for
asymmetric ciphers (\cite{Guan87}).

The rest of the paper is organized as follows: In Section $2$,
some basic concepts regarding memory cellular automata are
introduced. In Section $3$, the new secret sharing scheme for
b\&w, gray-level and color images is presented. The security of
the proposed scheme is analyzed and proved in Section $4$. In
Section $5$ several examples for different classes of images and
different schemes, are given. Finally, the conclusions of this
paper are presented in Section $6$.

\section{Memory Cellular Automata}

\emph{Two-dimensional finite cellular automata} (2D-CA for short)
are discrete dynamical systems formed by a finite two-dimensional
array of $r\times s$ identical objects called \emph{cells}, such
that each of them can assume a state. The state of each cell is an
element of the \emph{finite state set}, $S$. Throughout this paper
we will consider $S = \mathbb{Z}_{c}$, where $c = 2^{b}$ is the
number of colors of the image; \emph{i.e.}, if the image is a b\&w
image, then $b = 1$; for gray-level images the value is $b = 8$,
and if it is a color image, then $b = 24$.

The $\left( i,j\right) $-th cell is denoted by $\left\langle
i,j\right\rangle $, and the state of this cell at time $t$ is
$a_{ij}^{\left( t\right) }\in\mathbb{Z}_{c}$. The 2D-CA evolves
deterministically in discrete time steps, changing the states of
all cells according to a \emph{local transition function},
\[
f\colon\left( \mathbb{Z}_{c}\right)
^{n}\rightarrow\mathbb{Z}_{c}.
\]
The updated state of each cell depends on the $n$ variables of the
local transition function, which are the previous states of a set
of cells, including the cell itself, and constitute its
\emph{neighborhood}. For 2D-CA there are some classic types of
neighborhoods, but in this work only the \emph{extended Moore
neighborhood} will be consider; that is, the neighborhood of the
cell $\left\langle i,j\right\rangle $ is formed by its nine
nearest cells:
\begin{align*}
V_{i,j} & = \left\{ \left\langle i-1,j-1\right\rangle ,
\left\langle i-1, j\right\rangle ,\left\langle
i-1,j+1\right\rangle , \left\langle i,
j-1\right\rangle ,\right. \\
& \left. \left\langle i,j\right\rangle ,\left\langle
i,j+1\right\rangle , \left\langle i+1,j-1\right\rangle
,\left\langle i+1,j\right\rangle , \left\langle
i+1,j+1\right\rangle \right\} .
\end{align*}
Graphically it can be seen as follows:
\[
\begin{tabular}
[c]{|c|c|c|}\hline $\overset{\,}{\left\langle i-1,j-1\right\rangle
}$ & $\underset{\,} {\left\langle i-1,j\right\rangle }$ &
$\left\langle i-1,j+1\right\rangle $\\\hline
$\overset{\,}{\left\langle i,j-1\right\rangle }$ &
$\underset{\,}{\left\langle i,j\right\rangle }$ & $\left\langle
i,j+1\right\rangle $\\\hline $\overset{\,}{\left\langle
i+1,j-1\right\rangle }$ & $\underset{\,} {\left\langle
i+1,j\right\rangle }$ & $\left\langle i+1,j+1\right\rangle
$\\\hline
\end{tabular}
\]
Consequently the local transition function $f \colon\left(
\mathbb{Z} _{c}\right) ^{9}\rightarrow\mathbb{Z}_{c}$ is
\[
a_{ij}^{\left( t+1\right) }=f\left( a_{i-1,j-1}^{\left( t\right)
}, a_{i-1,j}^{\left( t\right) },a_{i-1,j+1}^{\left( t\right) },
a_{i,j-1}^{\left( t\right) },a_{ij}^{\left( t\right) },
a_{i,j+1}^{\left( t\right) },a_{i+1,j-1}^{\left( t\right) },
a_{i+1,j}^{\left( t\right) },a_{i+1,j+1}^{\left( t\right)
}\right),
\]
or equivalently,
\[
a_{ij}^{\left( t+1\right) }=f\left( V_{ij}^{\left( t\right)
}\right) , \quad0\leq i\leq r-1,\quad0\leq j\leq s-1,
\]
where $V_{ij}^{\left( t\right) }\subset\left(
\mathbb{Z}_{c}\right) ^{9}$ stands for the states of the neighbor
cells of $\left\langle i,j\right\rangle $ at time $t$. The matrix
\[
C^{\left( t\right) }=\left(
\begin{array}
[c]{ccc}
a_{00}^{\left( t\right) } & \cdots & a_{0,s-1}^{\left( t\right) }\\
\vdots & \ddots & \vdots\\
a_{r-1,0}^{\left( t\right) } & \cdots & a_{r-1,s-1}^{\left(
t\right) }
\end{array}
\right)
\]
is called the \emph{configuration at time} $t$ of the 2D-CA, and
$C^{\left( 0\right) }$ is the \emph{initial configuration} of the
CA. Moreover, the sequence $\{C^{\left( t\right) }\}_{0\leq t\leq
k}$ is called the \emph{evolution of order} $k$ of the 2D-CA, and
$\mathcal{C}$ is the set of all possible configurations of the
2D-CA; consequently $\left| \mathcal{C} \right| =c^{r\cdot s}$.

As the number of cells of the 2D-CA is finite, boundary conditions
must be considered in order to assure the well-defined dynamics of
the CA. In this paper, \emph{periodic boundary conditions} are
taken:
\[
a_{ij}^{\left( t\right) }=a_{uv}^{\left( t\right) }\Leftrightarrow
i\equiv u\text{ }\left( \operatorname{mod}r\right) ,\quad j\equiv
v\text{ } \left( \operatorname{mod}s\right) .
\]

The \emph{global function} of the CA is a linear transformation,
$\Phi \colon\mathcal{C}\rightarrow\mathcal{C}$, that yields the
configuration at the next time step during the evolution of the
CA, that is, $C^{\left( t+1\right) }=\Phi\left( C^{\left( t\right)
}\right) $. If $\Phi$ is bijective then there exists another
cellular automaton, called its \emph{inverse}, with global
function $\Phi^{-1}$. When such inverse cellular automaton exists,
the cellular automaton is called \emph{reversible} and the
evolution backwards is possible~(\cite{ToffoliMargolus90}).

Let us consider the set of 2D-CA whose local transition functions
are of the following form:
\[
a_{ij}^{\left( t+1\right) }=\sum_{\alpha,\beta\in\{-1,0,1\}}
\lambda _{\alpha,\beta}a_{i+\alpha,j+\beta}^{\left( t\right)
}\text{ } \left( \operatorname{mod}c\right) ,
\]
with $0\leq i\leq r-1$, $0\leq j\leq s-1$, and
$\lambda_{\alpha\beta} \in\mathbb{Z}_{2}$. They are called
2D-\emph{Moore} \emph{linear cellular automata }(2D-LCA). As there
are $9$ cells in the extended Moore neighborhood, then there exist
$2^{9}=512$ two-dimensional LCA, and every one of them can be
conveniently specified by a decimal integer called the \emph{rule
number}: $w$, which is defined as follows:
\begin{align*}
w & = \lambda_{-1,-1}2^{8}+\lambda_{-1,0}2^{7} +
\lambda_{-1,1}2^{6} + \lambda_{0,-1}2^{5} + \lambda_{0,0}2^{4}\\
& \quad + \lambda_{0,1}2^{3} + \lambda_{1,-1}2^{2} +
\lambda_{1,0}2^{1} + \lambda_ {1,1}2^{0},
\end{align*}
where $0\leq w\leq511$.

The standard paradigm for CA considers that the state of every
cell at time $t+1$ depends on the state of some cells (its
neighborhood) at time $t$. Nevertheless, one can consider CA for
which the state of every cell at time $t+1$ not only depends on
the states of some cells at time $t$, but also on the states of
(possible) other different groups of cells at times $t-1$, $t-2 $,
etc. This is the basic idea of \emph{memory cellular automata},
MCA for short, (see~\cite{Alonso03}). In this paper, we consider a
particular type of MCA called $k$\emph{-th order linear}
\emph{MCA} (LMCA for short) for which the local transition
function is of the following form:
\begin{equation}
a_{ij}^{\left( t+1\right) }=\sum_{m=0}^{k-1}f_{m+1} \left(
V_{ij}^{\left( t-m\right) }\right) \text{ } \left(
\operatorname{mod}c\right) ,\label{f3}
\end{equation}
with $0\leq i\leq r-1$, $0\leq j\leq s-1$, and where $f_{l}$,
$1\leq l\leq k$, are the local transition functions of $k$
particular 2D-LCA.

Note that the initial configuration of a 2D-LMCA is formed by $k$
components, $C^{\left( 0\right) },\ldots,C^{\left( k-1\right) }$,
in order to initialize the evolution of the MCA of order $k$.

A particular type of reversible MCA with local transition
function~(\ref{f3}) is characterized by means of the following
result.

\begin{proposition}
If the global function defining the 2D-CA with local transition
function $f_{k}$ is the identity, \emph{i.e.}, if
\[
f_{k}\left( V_{ij}^{\left( t-k+1\right) }\right) = a_{ij}^{\left(
t-k+1\right) },
\]
then the LMCA given by~\emph{(\ref{f3})} is a 2D-reversible MCA,
whose inverse CA is another LMCA with local transition function:
\begin{equation}
a_{ij}^{\left( t+1\right) } = -\sum_{m=0}^{k-2}f_{k-m-1} \left(
V_{ij}^{\left( t-m\right) }\right) + a_{ij}^{\left( t-k+1\right) }
\text{ }\left( \operatorname{mod}c\right) ,\label{f4}
\end{equation}
for $0\leq i\leq r-1$, $0\leq j\leq s-1$.
\end{proposition}

\begin{proof}
Suppose that $\{C^{\left( t\right) }\}_{t\geq0}$ is the evolution
of the LMCA given by~(\ref{f3}), where
\[
C^{\left( t\right) }=\left(
\begin{array}
[c]{ccc}
a_{00}^{\left( t\right) } & \cdots & a_{0,s-1}^{\left( t\right) }\\
\vdots & \ddots & \vdots\\
a_{r-1,0}^{\left( t\right) } & \cdots & a_{r-1,s-1}^{\left(
t\right) }
\end{array}
\right)
\]
is the configuration at time $t$, and let $\{\tilde{C}^{\left(
t\right) }\}_{t\geq0}$ be the evolution of the LMCA given
by~(\ref{f4}), where:
\[
\tilde{C}^{\left( t\right) }=\left(
\begin{array}
[c]{ccc}
\tilde{a}_{00}^{\left( t\right) } & \cdots &
\tilde{a}_{0,s-1}^{\left(
t\right) }\\
\vdots & \ddots & \vdots\\
\tilde{a}_{r-1,0}^{\left( t\right) } & \cdots &
\tilde{a}_{r-1,s-1}^{\left( t\right) }
\end{array}
\right) .
\]
The proof ends if we show that $\tilde{C}^{\left( k+1\right) } =
C^{\left( t-k+1\right) }$ when $\tilde{C}^{\left( 1\right) } =
C^{\left( t+1\right) }$, $\tilde{C}^{\left( 2\right) } = C^{\left(
t\right) }$, $\ldots$, $\tilde{C}^{\left( k\right) } = C^{\left(
t-k+2\right) }$, for every $t$. Consequently, by simply
applying~(\ref{f4}) we obtain:
\begin{equation}
\tilde{a}_{ij}^{\left( k+1\right) } = -f_{k-1} \left(
\tilde{V}_{ij}^{\left( k\right) }\right) - f_{k-2} \left(
\tilde{V}_{ij}^{\left( k-1\right) }\right) - \ldots - f_{1}\left(
\tilde{V}_{ij}^{\left( 2\right) }\right) + \tilde{a}_{ij}^{\left(
1\right) }\text{ } \left( \operatorname{mod} c\right) ,\label{f5}
\end{equation}
for $0\leq i\leq r-1,0\leq j\leq s-1$. As $\tilde{C}^{\left(
m\right) } = C^{\left( t-m+2\right) }$ with $1\leq m\leq k$, then
$\tilde{V} _{ij}^{\left( m\right) } = V_{ij}^{\left( t-m+2\right)
}$ with $1\leq m\leq k$. As a consequence, taking into account the
value of $C_{ij}^{\left( t+1\right) }$ given by~(\ref{f3}), the
equation~(\ref{f5}) yields:
\begin{align*}
\tilde{a}_{ij}^{\left( k+1\right) } & = -f_{k-1} \left(
V_{ij}^{\left( t-k+2\right) }\right) - \ldots - f_{1}\left(
V_{ij}^{\left( t\right)
}\right) + a_{ij}^{\left( t+1\right) }
\left( \operatorname{mod}c\right) \\
& = -f_{k-1}\left( V_{ij}^{\left( t-k+2\right) }\right) - \ldots
- f_{1}\left( V_{ij}^{\left( t\right) }\right) \\
& + f_{1}\left( V_{ij}^{\left( t\right) }\right) + \ldots +
f_{k-1}\left( V_{ij}^{\left( t-k+2\right) }\right) +
a_{ij}^{\left( t-k+1\right)
}\left( \operatorname{mod}c\right) \\
& = a_{ij}^{\left( t-k+1\right) }\text{ } \left(
\operatorname{mod} c\right) ,
\end{align*}
for every $0\leq i\leq r-1$, $0\leq j\leq s-1$, thus
$\tilde{C}^{\left( k+1\right) }=C^{\left( t-k+1\right) }$ and we
conclude.
\end{proof}

\section{A new graphic secret sharing scheme based on MCA}

In this section we propose a new graphic secret sharing scheme,
specifically a graphic $\left( k,n\right) $-threshold scheme,
based on memory cellular automata, in order to apply it to images.
Basically, it consists of considering the secret image as the
first component of the initial configuration for a 2D reversible
LMCA of order $k$, and the rest of $k-1$ components of the initial
configuration are $k-1$ random matrices. The shares to be
distributed among the $n$ participants are $n$ consecutive
configurations of the evolution of the LMCA. In the next
subsections, the scheme is more detailed.

\subsection{The representation of an image as a matrix}

An image $I$ defined by $c$ colors and $r\times s$ pixels,
$p_{ij}$, with $1\leq i\leq r$, $1\leq j\leq s$, can be considered
as a matrix $M$ with coefficients in $\mathbb{Z}_{c}$, as follows:

\begin{enumerate}
\item If $I$ is a b\&w image, then $M$ is an $r\times s$ matrix
whose $\left( i,j\right) $-th coefficient is $1$ (\emph{resp.
}$0$) if the pixel $p_{ij}$ is black (\emph{resp.} white);
\emph{i.e.}, the coefficients of $M$ are in $\mathbb{Z}_{2}$
($c=2$, and hence $b=1$).

\item If $I$ is a gray-level image, then the RGB code of each
pixel, $p_{ij} $, is given by the three-dimensional vector $\left(
R,G,B\right) $, where $0\leq R,G,B\leq255$ and $R = G = B$.
Consequently, each pixel can be defined by a number $0\leq
R\leq255$. Hence, $M$ is an $r\times s$ matrix with coefficients
in $\mathbb{Z}_{2^{8}}$.

\item Finally, if $I$ is a color image, then each pixel is given
by $24$ bits ($8$ bits representing each basic color: red, green
and blue). As a consequence $M$ is an $r\times s$ matrix with
coefficients in $\mathbb{Z} _{2^{24}}$.
\end{enumerate}

\subsection{The graphic sharing scheme}

As it is mentioned, the secret sharing scheme proposed is a
$\left( k,n\right) $-threshold scheme based on the use of a
2D-reversible LMCA. The protocol contains three phases: The setup
phase, the sharing phase and the recovery phase.

\subsubsection{The setup phase}

This first phase is given by the following steps:

\begin{enumerate}
\item The dealer generates a sequence of $k-1$ random integers
numbers:
\begin{equation}
\{w_{1},\ldots,w_{k-1}\}\label{omegas}
\end{equation}
such that $0\leq w_{l}\leq511$ with $1\leq l\leq k-1$. These
numbers stand for the rule numbers of the 2D-LCA constituting the
memory cellular automata.

\item The dealer constructs the 2D-LMCA with local transition
function:
\begin{equation}
a_{ij}^{\left( t+1\right) } = f_{w_{1}}\left( V_{ij}^{\left(
t\right) }\right) + \ldots + f_{w_{k-1}}\left( V_{ij}^{\left(
t-k+2\right) }\right) + a_{ij}^{\left( t-k+1\right) }\text{
}\left( \operatorname{mod}c\right) ,\label{ltf}
\end{equation}
where $f_{w_{l}} \colon\left( \mathbb{Z}_{c}\right) ^{9}
\rightarrow \mathbb{Z}_{c}$, and $0\leq i\leq r-1$, $0\leq j\leq
s-1$. The set of random numbers given in~(\ref{omegas}) should be
securely distributed to the participants if the dealer's role is
limited to elaborate the shares, and his help is not necessary to
recover the secret image.

\item The matrix representing the secret image to be shared is
considered as the first component of the initial configuration,
$C^{\left( 0\right) }=M$. Moreover, to complete the initial
configuration, the dealer generates $k-1$ random components:
$C^{\left( 1\right) },\ldots, C^{\left( k-1\right) }$, by means of
a cryptographic secure pseudorandom number generator
(see~\cite[Section 5.5]{MOV97}), in order to avoid an attack to
the scheme by supposing the values of these $k-1$ matrices. These
$k-1$ configurations must be destroyed after generating the
shares.
\end{enumerate}

\subsubsection{The sharing phase}

\begin{enumerate}
\item The dealer chooses a integer number $m$, such that $m\geq k$
in order to avoid possible overlaps between the initial conditions
and the shares. (Note that the number of iterations increases with
$m$, so this number would not be much bigger than $k$.)

\item Starting from the initial configurations $C^{\left( 0\right)
}, \ldots,C^{\left( k-1\right) }$, the dealer computes the $\left(
m+n-1\right) $-th order evolution of the 2D-LMCA:
\[
\left\{ C^{\left( 0\right) }, \ldots ,C^{\left( k-1\right) },
C^{\left( k\right) }, \ldots ,C^{\left( m\right) },\ldots,
C^{\left( m+n-1\right) }\right\} .
\]

\item The shares to be distributed among the $n$ participants,
$P_{0} ,\ldots,P_{n-1}$, are the last $n$ configurations computed:
$S_{0}=C^{\left( m\right) },\ldots,S_{n-1}=C^{\left( m+n-1\right)
}$. Moreover, each participant receives the set of random numbers
generated by the dealer in the step 1 of the setup phase, in order
to construct the inverse function of the local transition function
given by formula~(\ref{ltf}). In this way, each participant knows
how to recover the original image, without the cooperation of the
dealer.
\end{enumerate}

\subsubsection{The recovery phase}

To recover the secret image, any consecutive $k$ (of $n$) shared
images are needed, but no less. The following steps define this
phase.

\begin{enumerate}
\item To recover the secret, $C^{\left( 0\right) }$, a set of $k$
consecutive shares of the form
\[
S_{\alpha} = C^{\left( m+\alpha\right) },\ldots,S_{\alpha+k-1} =
C^{\left( m+\alpha+k-1\right) },\quad0\leq\alpha\leq n-k,
\]
is needed.

\item Taking $\tilde{S}_{0}=C^{\left( m+\alpha+k-1\right) },
\ldots , \tilde{S}_{k-1}=C^{\left( m+\alpha\right) }$, and
iterating $m+\alpha+k-1$ times the inverse LMCA, the secret
initial configuration (the original image), $C^{\left( 0\right)
}$, is obtained.
\end{enumerate}

Note that the recovered image is exactly the same than the
original one because the LMCA is reversible. This property of the
proposed scheme is not verified by any other graphic sharing
scheme. Moreover, as every participant knows the local transition
function, they do not need the collaboration of the dealer for
recovering the original image.

\section{Analysis of the security of the scheme}

In this section, the security of the proposed graphic sharing
scheme is analyzed. First of all, note that from the
formula~(\ref{rate}), the information rate of each participant of
this scheme is $1$, and consequently, the information rate for
this secret sharing scheme is also $1$. Furthermore, the scheme
proposed is ideal. It is also perfect, as it is proved in the
following

\begin{proposition}
Let us consider the LMCA given by the local transition
function~\emph{(\ref{f3})}. If one configuration of the form
$C^{\left( t-i\right) }$, $0\leq i\leq k-1$, is unknown, then no
information about the configuration $C^{\left( t+1\right) }$ can
be obtained.
\end{proposition}

\begin{proof}
Note that the evolution of the LMCA with local transition
function~(\ref{f3}) can be expressed in terms of global functions
as follows:
\begin{equation}
C^{\left( t+1\right) } = \Phi_{1}\left( C^{\left( t\right)
}\right) + \ldots + \Phi_{k-1}\left( C^{\left( t-k+2\right)
}\right) + \Phi_{k}\left( C^{\left( t-k+1\right) }\right) \text{ }
\left( \operatorname{mod} c\right) ,\label{f6}
\end{equation}
where $\Phi_{i}$ stands for the global function of the 2D-LCA with
transition function $f_{i}$. Now, without loss of generality, we
can assume that the unknown configuration is $C^{\left(
t-k+1\right) }$. Then the formula~(\ref{f6}) yields:
\[
C^{\left( t+1\right) }=U+V\text{ }\left(
\operatorname{mod}c\right) ,
\]
where $U=\left( u_{ij}\right) $ is a known matrix, and $V = \left(
v_{ij}\right) $ is the unknown matrix $\Phi_{k}\left( C^{\left(
t-k+1\right) }\right) $, where $0\leq i\leq r-1$, $0\leq j\leq
s-1$. Both matrices have coefficients in $\mathbb{Z}_{c}$. As a
consequence, the following linear system holds:
\[
a_{ij}^{\left( t+1\right) } = u_{ij}+v_{ij}\text{ } \left(
\operatorname{mod} c\right) ,\quad 0 \leq i\leq r-1,\quad0\leq
j\leq s-1,
\]
which is formed by $r\cdot s$ equations with $2r\cdot s$ unknowns.
Consequently, it can not be solved and, obviously, no information
about the configuration
\[
C^{\left( t+1\right) } = \left( a_{ij}^{\left( t+1\right) }\right)
,\quad 0 \leq i\leq r-1, \quad0\leq j\leq s-1,
\]
is obtained.
\end{proof}

Remark that a similar result holds if the number of unknown
configurations is greater than one. As a consequence, for the
secret sharing scheme proposed it is impossible to recover the
secret image from $k-1$ (or less) shares.

Furthermore, it is assumed that each participant knows the local
transition function, but this knowledge does not suppose a
weakness of the scheme and it permits to recover the secret image
without the collaboration of the dealer.

\section{An example}

In this section, an example for a $(k,n)$-threshold
scheme will be presented.
The random matrices (configurations)
$C^{(1)},C^{(2)},\ldots,C^{(k-1)}$, for the step $3 $ in the setup
phase have been generated by using the BBS pseudorandom bit
generator~(\cite{BBS86}). This generator is a cryptographically
secure pseudorandom bit generator and it is defined by iterating
the function $x^{2}\left( \operatorname{mod}n\right) $, where
$n=p\cdot q$ is the product of two large prime numbers, each of
them congruent to $3$ modulo $4$. In other words, the BBS
generator produces a sequence of bits by taking the least
significant bit of the sequence defined by $x_{i+1}\equiv
x_{i}^{2}\left( \operatorname{mod}n\right) $, $i\geq0$, where
$x_{0}$ is the seed of the generator. The conditions for choosing
the modulus $n$ and the seed $x_{0}$, in order to obtain orbits of
maximal periods have been established in~\cite{HMMP98}.

In the practical implementation of the proposed scheme for the
following example, we have decided to obtain shared images of the
same type than the original one, that is, as the original image
is a gray-level image, the shares also will be gray-level images.
An easier implementation could determine shares with $2^{24} $ colors, in
all cases, \emph{i.e.}, without taking into account the number of
colors of the secret image.

\begin{figure}[h]
\begin{center}
\includegraphics{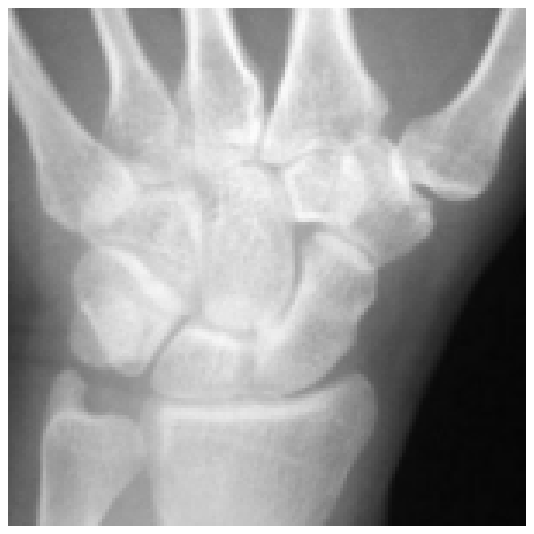}
\end{center}
\caption{Secret image} \label{fig:grey}
\end{figure}

\begin{figure}[h]
\begin{center}
\includegraphics{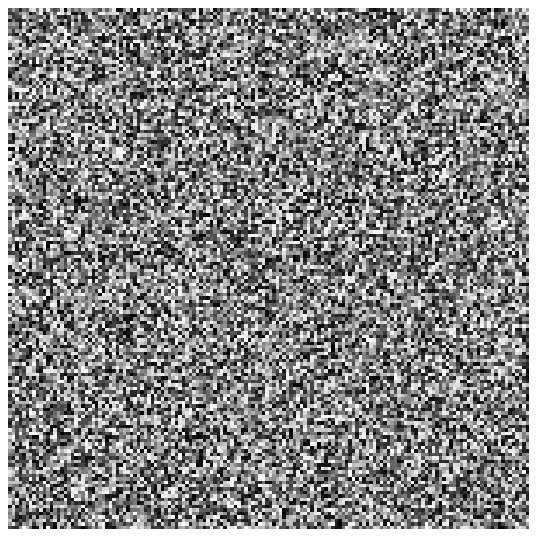}
\end{center}
\caption{First share} \label{fig:share0}
\end{figure}

\begin{figure}[h]
\begin{center}
\includegraphics{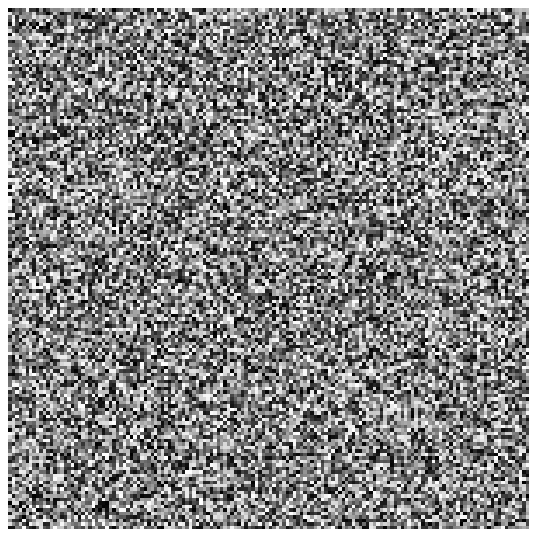}
\end{center}
\caption{Second share} \label{fig:share1}
\end{figure}

\begin{figure}[h]
\begin{center}
\includegraphics{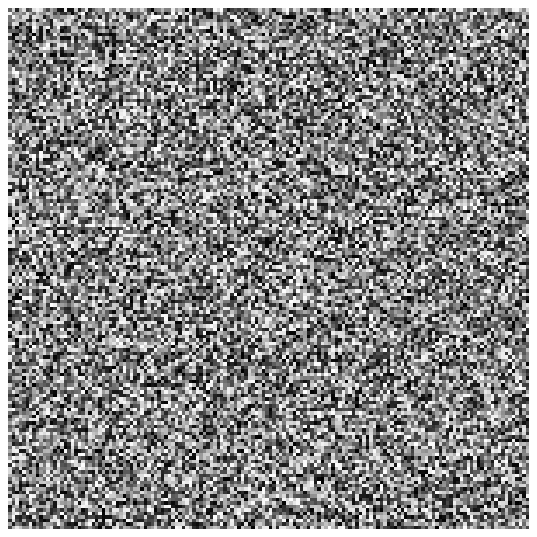}
\end{center}
\caption{Third share} \label{fig:share2}
\end{figure}

\begin{figure}[h]
\begin{center}
\includegraphics{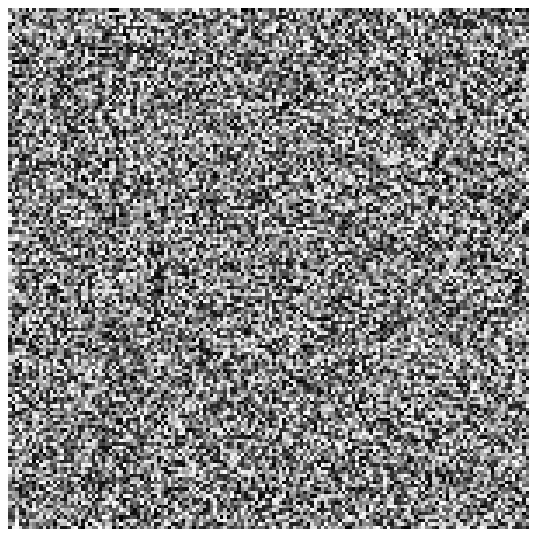}
\end{center}
\caption{Fourth share} \label{fig:share3}
\end{figure}

\begin{figure}[h]
\begin{center}
\includegraphics{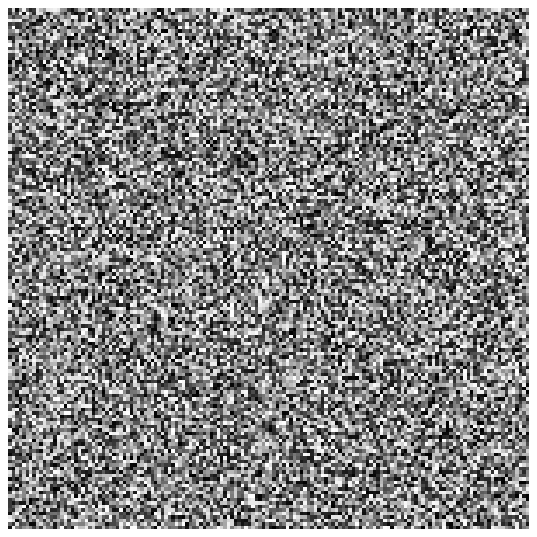}
\end{center}
\caption{Fifth share} \label{fig:share4}
\end{figure}

The image used in the following example has
$181\times157$ pixels and $249$ gray-levels (see Figure~\ref{fig:grey}).
Moreover, the local function used is
defined by means of the number $w_{1} = 232$, and the parameters
are:
\[
k = 3,\quad n = 5,\quad m = 3.
\]
In this way, $5$ iterations for the CA have been needed in order
to obtain the following images:
\[
C^{(0)}, C^{(1)}, C^{(2)}, S_{0} = C^{(3)}, S_{1} = C^{(4)},
S_{2} = C^{(5)}, S_{3} = C^{(6)}, S_{4} = C^{(7)}.
\]
The five shares are shown as images in Figures \ref{fig:share0},
\ref{fig:share1}, \ref{fig:share2}, \ref{fig:share3}, and
\ref{fig:share4}, respectively.

For this example we have used the shares
$S_{2}$, $S_{3}$ and $S_{4}$ to recover, after $5$ iterations, the
original image (see Figure~\ref{fig:grey}).

\section{Conclusions}

In this paper a new graphic $\left( k,n\right)$-threshold scheme
for sharing secret b\&w, gray-level and color images is presented.
The scheme is based on two-dimensional reversible linear memory
cellular automata. The two main characteristics of this new
scheme, which are not satisfied by any previous proposed graphic
schemes, are:\ (1) The size of each shared image is exactly the
same than the size of the secret image to be shared, and (2) There
is no loss of resolution in the recovery secret image.

Moreover, the security of the scheme has been analyzed and it has
been proved that in order to obtain the original image it is
necessary to join, at least, $k$ shares. If $k-1$ or less shares
are pooled, no information of the secret image is obtained.

\section*{Acknowledgments}

This work is supported by Ministerio de Ciencia y Tecnolog\'{i}a
(Spain), under grant TIC2001-0586, and by the Consejer\'{i}a de
Educaci\'{o}n y Cultura of Junta de Castilla y Le\'{o}n (Spain),
under grant SA052/03.

\newpage

\section*{Figure captions}

\end{document}